\documentclass[aps,prl,reprint,amsmath,amssymb,superscriptaddress,nobibnotes]{revtex4-1}
\usepackage{graphicx,SIunits}
\usepackage{bm}     
\usepackage{hyperref} 
\usepackage{dsfont}    
\usepackage{color}

\usepackage{multirow}

\begin{document}
	
\title{Distributed quantum sensing with multi-mode $N00N$ states}
	
\author{Dong-Hyun Kim}
\email{These authors contributed equally} 
\affiliation{Center for Quantum Technology, Korea Institute of Science and Technology (KIST), Seoul, 02792, Korea}
\affiliation{Department of Physics, Yonsei University, Seoul 03722, Korea}
	
\author{Seongjin Hong}
\email{These authors contributed equally}
\affiliation{Department of Physics, Yonsei University, Seoul 03722, Korea}

\author{Yong-Su Kim}
\affiliation{Center for Quantum Technology, Korea Institute of Science and Technology (KIST), Seoul, 02792, Korea}
\affiliation{Division of Quantum Information, KIST School, Korea University of Science and Technology, Seoul 02792, Korea}
	
\author{Kyunghwan Oh}
\affiliation{Department of Physics, Yonsei University, Seoul 03722, Korea}
	
\author{Su-Yong Lee}
\affiliation{Emerging Science and Technology Directorate, Agency for Defense Development, Daejeon 34186, Korea}
\affiliation{Weapon Systems Engineering, ADD School, University of Science and Technology, Daejeon, 34060, Korea}
	
\author{Changhyoup Lee}
\affiliation{Korea Research Institute of Standards and Science, Daejeon 34113, Korea}
	
\author{Hyang-Tag Lim}
\email{hyangtag.lim@kist.re.kr}
\affiliation{Center for Quantum Technology, Korea Institute of Science and Technology (KIST), Seoul, 02792, Korea}
\affiliation{Division of Quantum Information, KIST School, Korea University of Science and Technology, Seoul 02792, Korea}
	
\date{\today} 
	
\begin{abstract}
Distributed quantum sensing, which estimates a global parameter across distant nodes, has attracted significant interest for applications such as quantum imaging, sensor networks, and global-scale clock synchronization. $N00N$ states are regarded as one of the optimal quantum resources for quantum metrology, enabling the Heisenberg scaling. Recently, the concept of $N00N$ states has been extended to multi-mode $N00N$ states for quantum-enhanced multiple-parameter estimation. However, the application of multi-mode $N00N$ states in distributed quantum sensing remains unexplored. Here, we propose a distributed quantum sensing scheme that achieves the Heisenberg scaling using multi-mode $N00N$ states. We theoretically show that multi-mode $N00N$ states can reach the Heisenberg scaling by examining both the Cram\'er-Rao bound and the quantum Cram\'er-Rao bound. For experimental demonstration, we employ a four-mode $2002$ state to estimate the average of two spatially distributed phases, achieving a 2.74 dB sensitivity enhancement over the standard quantum limit. We believe that utilizing multi-mode $N00N$ states for distributed quantum sensing offers a promising approach for developing entanglement-enhanced sensor networks.
\end{abstract}
\maketitle
	
Exploiting quantum resources, such as entanglement and squeezing, can enhance sensitivity in estimating unknown parameters, surpassing the standard quantum limit (SQL) imposed by classical resources~\cite{dowling2008,giovannetti2011,pirandola2018,lawrie2019}. In single-parameter estimation, it is well known that $N00N$ states can achieve the Heisenberg scaling (HS) of $1/N^2$ due to their photon-number-path entanglement~\cite{polino2020,grun2022}. Distributed quantum sensing, in particular, has gained significant attention for its broad potential applications, including quantum imaging~\cite{brida2010,hugo2019}, phase tracking~\cite{hidehiro2012}, global clock synchronization~\cite{komar2014,malia2022}, dark matter searches~\cite{backes2021,shi2023,brady2023}, and sensor networks~\cite{eldredge2018,gessner2020}.	
	
The goal of distributed quantum sensing is to estimate a global parameter from spatially distributed unknown parameters with quantum-enhanced sensitivity~\cite{ge2018,proctor2018,oh2020,zhang2021,oh2022,liu2024}. Numerous studies have shown that the HS can be achieved by leveraging quantum resources, particularly entanglement, in distributed quantum sensing~\cite{guo2020,hong2024, liu2021, zhao2021, kim2024}. For example, multi-mode squeezed entangled states~\cite{zhuang2020,guo2020,hong2024,nehra2024} can achieve the HS in continuous-variable systems. Additionally, mode-entangled spin-squeezed atomic states~\cite{malia2022} and mode- and particle-entangled (MePe) states~\cite{gessner2018,liu2021,zhao2021,kim2024} enable the HS in discrete-variable systems.
	
Recently, the concept of $N00N$ states has been extended to generalized multi-mode $N00N$ states~\cite{humphreys2013}. The entanglement among multiple modes allows multi-mode $N00N$ states to serve as a promising candidate for probe states in multiple-parameter estimation~\cite{humphreys2013,hong2021}. Both theoretical and experimental studies have demonstrated that generalized multi-mode $N00N$ states can achieve enhanced sensitivity, outperforming other quantum states in simultaneous multiple-parameter estimation~\cite{humphreys2013,hong2021,hong2022,rehman2022,namkung2024}. This multi-mode entanglement in multi-mode $N00N$ states provides a natural approach to implementing distributed quantum sensing. However, despite recent advancements, multi-mode $N00N$ states have so far been applied only to measure simultaneous multiple phases~\cite{hong2021,hong2022,namkung2024,namkung20242}, and their potential for distributed quantum sensing remains unexplored.
	
In this work, we theoretically analyze the Cram\'er-Rao bound (CRB) and the quantum Cram\'er-Rao bound (QCRB) of the multi-mode $N00N$ states. We also compare the sensitivity gain of multi-mode $N00N$ states with other probe states, including separable $N00N$ states and MePe states~\cite{gessner2018, liu2021}. Furthermore, we experimentally estimate the average of the two unknown phases using multi-mode $N00N$ states with photon number $N=2$ and mode number $4$ in a distributed quantum sensing setup. By employing the maximum likelihood estimator (MLE), we show that the sensitivity achieves the HS, surpassing the SQL. Our results offer valuable insights for advancing distributed quantum sensor networks with quantum-enhanced sensitivity.

\begin{figure}[t]
\includegraphics[width=\columnwidth]{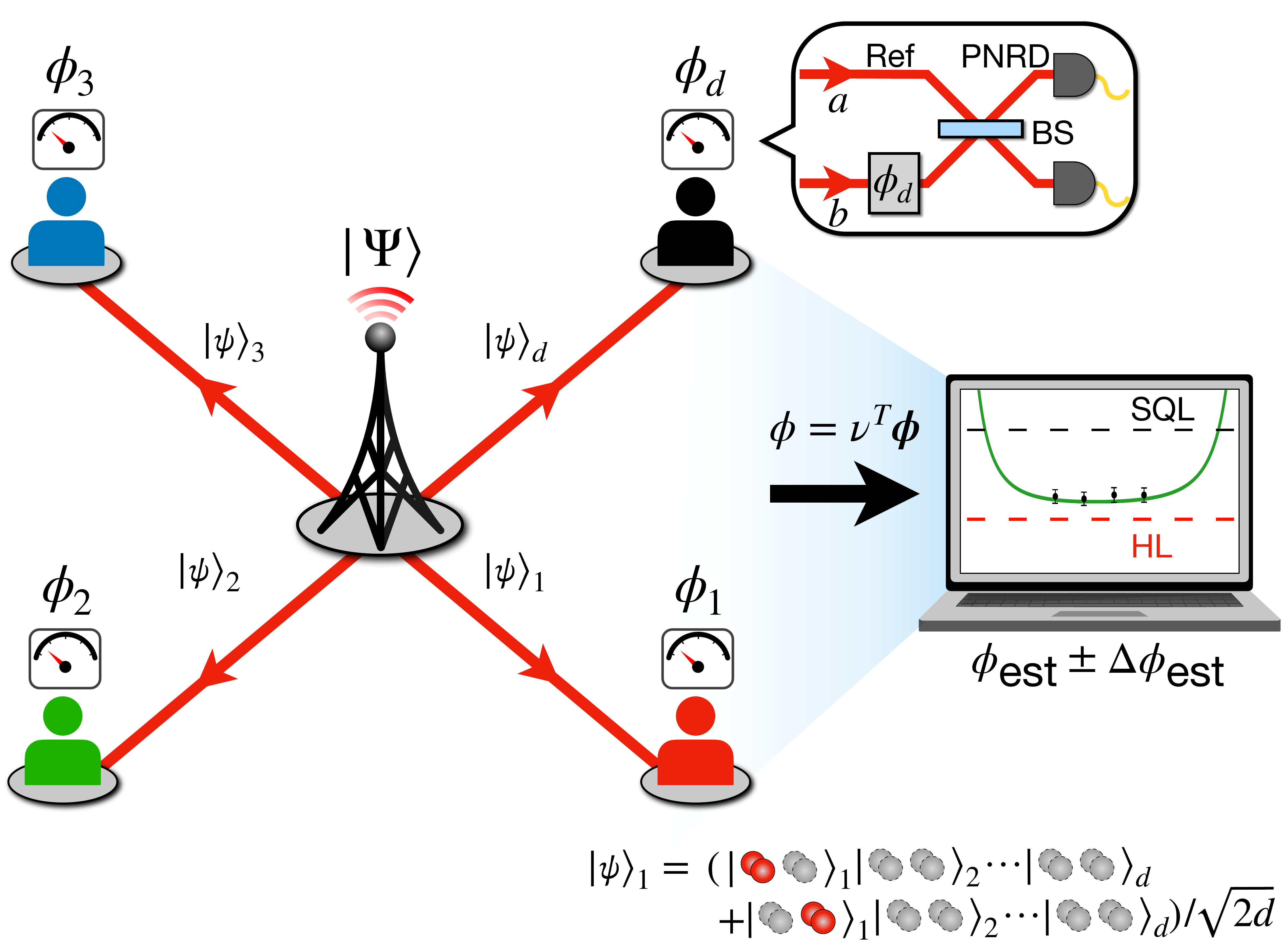} 
\caption{{\bf Schematic of distributed quantum sensing for estimating $d$ spatially distributed phases.} Probe states $|\Psi \rangle$ are sent to each node to estimate the distributed multiple phases $\{\phi_j\}$. The unknown phases to be measured are the phase differences between the arms of interferometers at each node. The probe state $|\psi\rangle_j$ undergoes the individual phase shift on each node and is then measured by local measurement. This enables the estimation of a linear combination of unknown phases with quantum-enhanced sensitivity.}
\label{fig:figure1}
\end{figure}
	
\begin{figure*}[t]
\includegraphics[width=\textwidth]{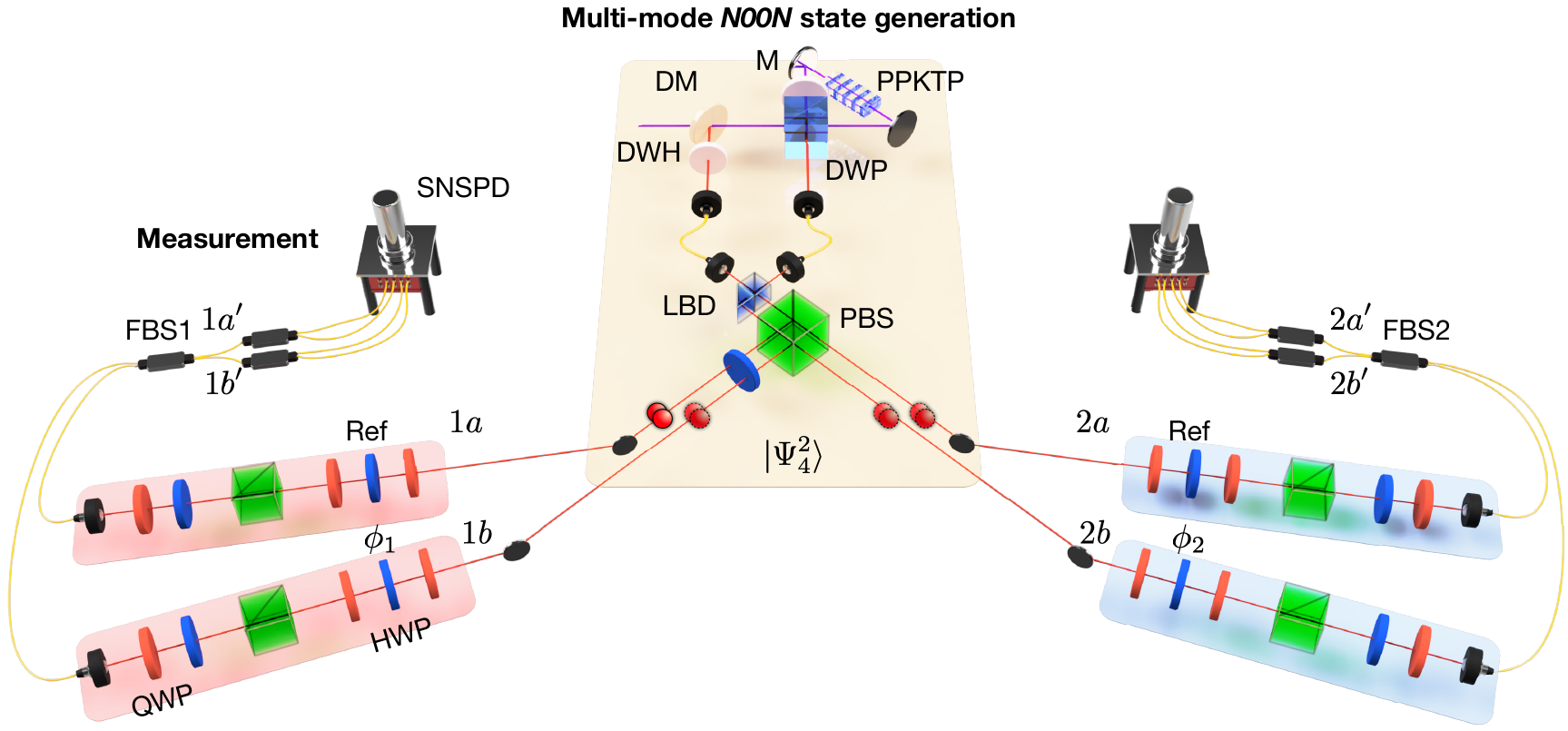} 
\caption{{\bf Experimental setup for estimating the average of two phases.} The four-mode $2002$ state $\left |\Psi_4^2 \right \rangle$  is generated using the Bell states and the Hong-Ou-Mandel interference~\cite{hong2021}, and then distributed to each node, where it undergoes phase shifts using a combination of two QWPs, and a HWP at each node. Then, after passing through FBS1(2), the probe state in modes 1$a$ and 1$b$ (2$a$ and 2$b$) is transmitted to output ports of 1$a'$ and $1b'$ (2$a'$ and $2b'$). Then, we measure the number of photons at each output port using a pseudo-photon number resolving detector (PNRD), consisting of FBSs and superconducting nanowire single-photon detectors (SNSPDs). QWP: quarter waveplate; HWP: half waveplate; LBD: lateral beam displacer; PBS: polarizing beam splitter; DM: dichroic mirror; DWP: dual wavelength PBS; M: mirror; DWH: dual wavelength HWP; PPKTP: periodically-poled KTiOPO; FBS: $50/50$ fiber beam splitter.}
\label{fig:figure2}
\end{figure*}
	
	
	
 Let us begin by introducing a general scenario of distributed quantum sensing for estimating the spatially distributed $d$ phases, as shown in Fig.~\ref{fig:figure1}~\cite{proctor2018, gessner2018, liu2021}. There are $d$ unknown phases, $\bm{\phi}=(\phi_1,\phi_2,...,\phi_d)$, each located at different locations as shown in Fig.~\ref{fig:figure1}. In distributed quantum sensing, our goal is to estimate a linear global function of these phases, given by $\phi=\bm{\nu}^T \bm{\phi}$, where $\bm{\nu}=(\nu_1, \nu_2, \dots,\nu_d)$ and $\sum_{j=1}^{d}{|\nu_j|}=1$~\cite{proctor2018, gessner2018, liu2021}. We define the $d$ unknown phases as the phase differences between the two arms of an interferometer at each node, setting $\nu_j=1/d$ for all $j$. In this scenario, we have $d$ nodes with two modes per interferometer, leading to a total of $2d$ modes. The initial probe state $|\Psi\rangle$ undergoes phase encoding $\hat{U}(\bm{\phi})$ and evolved into $\hat{U}(\bm{\phi})|\Psi\rangle$. Details on the phase encoding are provided in Supplemental Meterial~\cite{supple}. The encoded state is then detected by a set of projectors $\{\hat{\Pi}_l\}$, yielding a corresponding probability set of $\{P_l\}$. Finally, we can estimate the $\phi_\text{est}$ from $\{P_l\}$ using proper estimators~\cite{gessner2018,kim2024}. 
	
Here the uncertainty bound for the estimation of $\phi$ can be described as follows:
\begin{equation}
	\Delta^2\phi\equiv \langle(\phi_\text{est}-\phi)^2\rangle \geq \dfrac{\bm{\nu}^T \bold{F}_\text{C}^{-1} \bm{\nu}}{\mu} \geq \dfrac{\bm{\nu}^T \bold{F}_\text{Q}^{-1} \bm{\nu}}{\mu}.
	\label{eq:SB}
\end{equation}
Their bounds are called CRB and QCRB, respectively~\cite{gessner2018}. $\bold{F}_\text{C}$ denotes the classical Fisher information matrix (CFIM), obtained by using 	$F_{\text{C}(j,k)}=\sum_{l}(1/P_l)(\partial P_l / \partial \phi_j)(\partial P_l / \partial \phi_k)$, and the quantum Fisher information matrix (QFIM) can be defined by $F_{\text{Q}(j,k)}=4\text{Re}[\langle\partial_{\phi_j} \Psi|\partial_{\phi_k} \Psi\rangle - \langle \partial_{\phi_j} \Psi|\Psi\rangle\langle \Psi|\partial_{\phi_k}\Psi\rangle]$. In Eq.~(\ref{eq:SB}), $\mu$ denotes the number of measurements, and we set $\mu=1$ to simplify the notation in the theoretical schemes.

		
In our work, we consider the multi-mode $N00N$ states, $|\Psi_\text{MN}\rangle$, as a probe state, which has the form of 
\begin{equation}
	|\Psi_\text{MN}\rangle =  \dfrac{1}{\sqrt{d}}\sum_{j=1}^{d}(\dfrac{1}{\sqrt{2}}(|N0\rangle_{j}+|0N\rangle_{j})\bigotimes_{j\neq k}^{d}|00\rangle_{k}).
	\label{eq:MN}
\end{equation}
$|\Psi_\text{MN}\rangle$ is a coherent superposition of all configurations in which $N$ photons are present in one mode and none in any of the other $2d-1$ modes~\cite{hong2021}. The multi-mode $N00N$ states, in Eq.~(\ref{eq:MN}), can also be described as $|\Psi_\text{MN}\rangle = \sum_{j=1}^{d} |\psi\rangle_j$, where, for example, $|\psi\rangle_1=(|N0\rangle_1|00\rangle_2\dots|00\rangle_d+|0N\rangle_1|00\rangle_2\dots|00\rangle_d)/\sqrt{2d}$ and $|N0\rangle_1|00\rangle_2=|N_{1a}0_{1b}\rangle \bigotimes |0_{2a}0_{2b}\rangle$, where $a$ and $b$ denote two modes in the interferometer at each node as shown in Fig.~\ref{fig:figure1}. $|\Psi_\text{MN}\rangle$ can be prepared at the central node, with each $|\psi\rangle_j$ distributed to  the corresponding node, as shown in Fig.~\ref{fig:figure1}.
	
Then, we investigate the sensitivity bound for $|\Psi_\text{MN}\rangle$ and compare it with separable two-mode $N00N$ states, which are products of multiple copies of two-mode $N00N$ states~\cite{hong2021}. The separable two-mode $N00N$ state, $|\Psi_{\text{SN}}\rangle$, has the form of 
\begin{equation}
	|\Psi_{\text{SN}}\rangle = \bigotimes_{j=1}^{d} \dfrac{1}{\sqrt{2}}(|\dfrac{N}{d}0\rangle_{j}+|0\dfrac{N}{d}\rangle_{j}).
	\label{eq:SN}
\end{equation}

Note that $|\Psi_{\text{SN}}\rangle$ is the product state of $N/d$-photon two-mode $N00N$ states, while $|\Psi_\text{MN}\rangle$ represents the $N$-photon multi-mode states. For the practical realization of $|\Psi_{\text{SN}}\rangle$, $N/d$ is expected to be a positive integer. However, for convenience, we consider all possible cases of $N/d$ here. Thus, the total photon numbers are the same in both $|\Psi_{\text{SN}}\rangle$ and $|\Psi_\text{MN}\rangle$. We analyze the fundamental sensitivity bound governed by the QCRB for estimating $\phi$ using Eq~.(\ref{eq:SB}) with $|\Psi_\text{MN}\rangle$ and $|\Psi_\text{SN}\rangle$, respectively. Moreover, we calculate the CRB by considering local measurements at each node.
	
First, we analyze the sensitivity bound for the multi-mode $N00N$ states, $|\Psi_\text{MN}\rangle$. After undergoing phase encoding, $|\Psi_\text{MN}\rangle$ becomes as $\hat{U}(\bm{\phi})|\Psi_\text{MN}\rangle$. The QFIM of $|\Psi_\text{MN}\rangle$ can then be calculated to be
\begin{eqnarray}
	\bold{F}^\text{MN}_\text{Q}=\left(\begin{array}{cccc}
	(2d-1)N^2/d^2 & -(N/d)^2 & \cdots & -(N/d)^2 \\
	 -(N/d)^2 &   &  &  \vdots \\
	\vdots & \;  & \ddots &  -(N/d)^2 \\
	-(N/d)^2 &  \cdots & -(N/d)^2 & (2d-1)N^2/d^2
\end{array}  \right).
\label{eq:MNQF}
\end{eqnarray}
Note that all diagonal terms are $(2d-1)N^2/d^2$ and all off-diagonal terms are $-(N/d)^2$. The QCRB of $|\Psi_\text{MN}\rangle$ can then be obtained as 1/$N^2$ using Eq.~(\ref{eq:SB}) with $\bm{\nu}=(1/d, 1/d, ...,1/d)$. One can find that $|\Psi_\text{MN}\rangle$ can achieve the HS~\cite{gessner2018, liu2021, kim2024}. Then, we analyze the CRB, which serves as the practical sensitivity by considering the measurement. Here, we take into account local measurements consisting of a $2 \times 2$ beam splitter (BS) and a photon number-resolving detector (PNRD) at each node, as shown in the upper right inset of Fig.~\ref{fig:figure1}. The CFIM of $|\Psi_{\text{MN}}\rangle$, $\bold{F}_\text{MN}$, can be described as follows:
\begin{eqnarray}
	\bold{F}_\text{C}^\text{MN}=\left(\begin{array}{cccc}
		N^2/d & 0 & \cdots & 0 \\
		0 &   &   & \vdots \\
		\vdots & \; & \ddots & 0 \\
		0 & \cdots & 0 & N^2/d
	\end{array}  \right).
	\label{eq:MNCF}
\end{eqnarray}
See Supplemental Material for the calculation of CFIM~\cite{supple}.

In $\bold{F}_\text{C}^\text{MN}$, all diagonal terms are $N^2/d$, and all off-diagonal terms are 0. We then calculate the sensitivity bound of $|\Psi_\text{MN}\rangle$ using Eq.~(\ref{eq:SB}) as follows:
\begin{equation}
	\Delta^2\phi_\text{MN} \geq \dfrac{1}{N^2}.
	\label{eq:MNB}
\end{equation}
Our results clearly show that the HS can be achieved with $|\Psi_\text{MN}\rangle$ and optimal local measurement consisting of BS and PNRDs. It is noteworthy that, in distributed quantum sensing, the use of local measurement provides an advantage for estimating the spatially distributed parameters when sensing nodes are delocalized~\cite{zhao2021,malitesta2021}.
	
Then, we calculate the sensitivity bounds of separable two-mode $N00N$ states, i.e., $|\Psi_\text{SN}\rangle$ from Eq.~(\ref{eq:SN}), and compare it to $|\Psi_\text{MN}\rangle$. The corresponding QFIM of $|\Psi_{\text{SN}}\rangle$ is obtained as
\begin{eqnarray}
	\bold{F}_{\text{Q}}^\text{SN}=\left(\begin{array}{cccc}
		(N/d)^2 & 0 & \cdots & 0 \\
		0 &   &   & \vdots \\
		\vdots & \; & \ddots & 0 \\
		0 & \cdots & 0 & (N/d)^2
	\end{array}  \right).
	\label{eq:SNQF}
\end{eqnarray}
In $\bold{F}^{\text{SN}}_\text{Q}$, all diagonal terms are $(N/d)^2$, and all off-diagonal terms are zero. The QCRB of $|\Psi_\text{SN}\rangle$ is calculated to be $d/{N^2}$, which provides a worse sensitivity bound compared to $1/N^2$ of $|\Psi_\text{MN}\rangle$. The CRB of $|\Psi_\text{SN}\rangle$ can be achieved at the QCRB through optimal measurements involving a 2$\times$2 BS and PNRDs, thus yielding the CRB of $d/{N^2}$~\cite{humphreys2013}. The sensitivity bound of separable $N00N$ states for estimating $\phi$ is obtained as follows:
\begin{equation}
	\Delta^2\phi_\text{SN} \geq \dfrac{d}{N^2}.
	\label{eq:MNB}
\end{equation}
See Supplemental Material for detailed calculations~\cite{supple}.

Then, we analyze the figure of merit for the sensitivity of three probe states: $|\Psi_\text{SN}\rangle$, $|\Psi_\text{MN}\rangle$, and $|\Psi_\text{MePe}\rangle$, which is the MePe states that achieve the best sensitivity of $1/N^2$~\cite{gessner2018, liu2021}. See Supplemental Material for detailed calculations.~\cite{supple}. To this end, we define the gain parameter $G$ in sensitivity of probe states relative to the SQL as follows: $G=\Delta^2\phi_\text{SQL}/\Delta^2\phi$, where $\Delta^2\phi_\text{SQL}=1/N$ is the sensitivity bound of product states of multiple coherent states. The gain and sensitivity bounds of the SQL, $|\Psi_{\text{SN}}\rangle$, $|\Psi_{\text{MN}}\rangle$, and  $|\Psi_\text{MePe}\rangle$ are summarized in Table~\ref{table:CRB}. Here, we find that $|\Psi_\text{MN}\rangle$ with local measurements can achieve the HS equivalent to MePe states, which is known to be the best sensitivity bound~\cite{gessner2018, liu2021}. Moreover, $|\Psi_{\text{SN}}\rangle$ gives a worse sensitivity than the SQL when the number of unknown phases is larger than the total number of photons, i.e., $d>N$.
\begin{table}[h]
	\centering
	\begin{tabular}{c c c }
		\hline
		\hline
		Probe state & Sensitivity bound ($\Delta^2\phi$) & Gain ($G$)  \\
		\hline
		Classical state & $1/N$ & 1 \\
		$|\Psi_{\text{SN}}\rangle$ & $d/N^2$  & $N/d$ \\
		$|\Psi_{\text{MN}}\rangle$ & $1/N^2$  & $N$ \\
		$|\Psi_{\text{MePe}}\rangle$~\cite{gessner2018,liu2021} & $1/N^2$ & $N$   \\
		\hline
		\hline
	\end{tabular}
	\caption{\bf{Sensitivity bound, $\Delta^2\phi$, and gain, $G$, for the different probe states.}}
	\label{table:CRB}
\end{table}

Our experimental setup is depicted in Fig.~\ref{fig:figure2}. The procedure to prepare the multi-mode $N00N$ states for distributed quantum sensing is the following~\cite{hong2021}:

\begin{eqnarray}
	\left |\Phi^+ \right \rangle&=&(|H H\rangle+|V V\rangle)/\sqrt{2}\nonumber
	\\   
	\xrightarrow{{\rm LBD, PBS, HWP}}
	|\Psi_4^2\rangle&=&(|20\rangle_{1}|00\rangle_{2}+|02\rangle_{1}|00\rangle_{2}\nonumber
	\\
	&+&|00\rangle_{1}|20\rangle_{2}+|00\rangle_{1}|02
	\rangle_{2})/2,\nonumber
	\\
	\label{eq:MNN}
\end{eqnarray}
where subscript 1 (2) denotes the nodes and $|20\rangle_1|00\rangle_2=|2_{1a}0_{1b}\rangle \bigotimes |0_{2a}0_{2b}\rangle$. See Supplemental Material for detailed information on generating the multi-mode $N00N$ state~\cite{supple}. Then, our probe state $|\Psi_4^2\rangle$ is sent to each node and undergoes a phase shift $\{\phi_{1},\phi_{2}\}$, implemented via the combination of two quarter waveplates (QWPs) at optical axis 45$\degree$ and the half waveplate (HWP) as shown in Fig.~\ref{fig:figure2}. The output states are measured using the 50/50 fiber beam splitters (FBS) and superconducting nanowire single-photon detectors (SNSPDs) at each node. Note that both output ports of the FBS1 (FBS2) are connected to the two FBSs to implement a pseudo-PNRD. Then, we can obtain a post-selected two-photon detection probability set of $\{P_l\}=\{P_{1}^{11}$, $P_{1}^{20}$, $P_{1}^{02}$, $P_{2}^{11}$, $P_{2}^{20}$, $P_{2}^{02}\}$ with positive operator-valued measures (POVM) $\{\hat{\Pi}_{l}\}=\{|20\rangle\langle20|_{1}, |11\rangle\langle11|_{1}, |02\rangle\langle02|_{1}, |20\rangle\langle20|_{2}, |11\rangle\langle11|_{2}, |02\rangle\langle02|_{2}\}$. Here, $|20\rangle\langle20|_{1(2)}$ denotes the projectors for measuring two photons in node $1a'\;(2a')$ and no photons in node $1b'\;(2b')$, while $|11\rangle\langle11|_{1(2)}$ represents the projector for measuring one photon in node $1a'\;(2a')$ and one photon in node $1b'\;(2b')$, as shown in Fig.~\ref{fig:figure2}. See Supplemental Material for detailed information on the detection probability~\cite{supple}. Then, the sensitivity bound can be calculated as $1/4$ using Eq.~(\ref{eq:SB}), which means the HS can be achieved by using four-mode $2002$ states beyond the SQL.

\begin{figure}[t]
\includegraphics[width=\columnwidth]{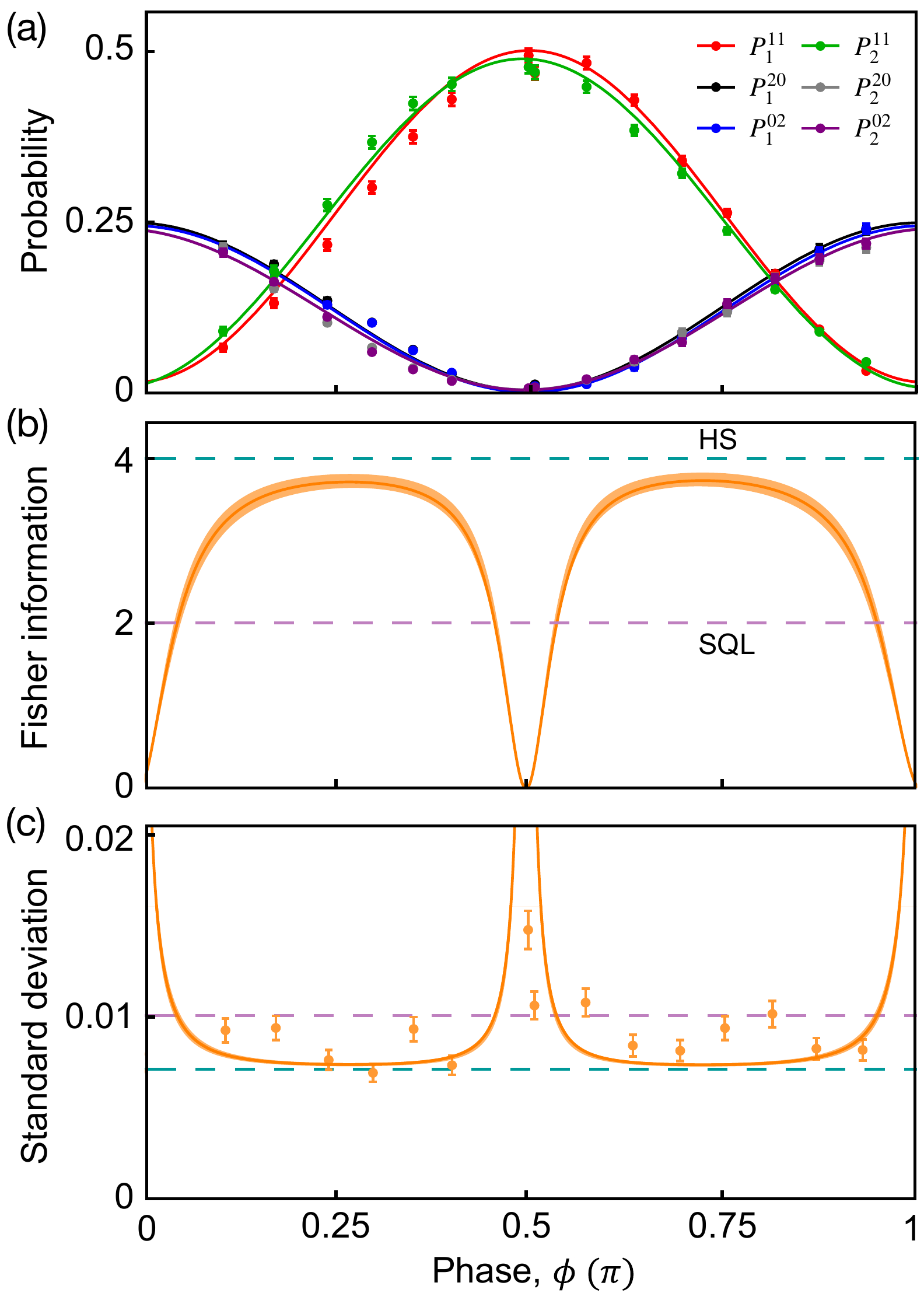}
\caption{{\bf Experimental results on the detection probabilities, the Fisher information, and standard deviation on the estimated phase $\phi_\text{est}$.} (a) Experimentally measured probabilities of $P_{1}^{11}$, $P_{1}^{20}$, $P_{1}^{02}$, $P_{2}^{11}$, $P_{2}^{20}$, and $P_{2}^{02}$. Solid lines represent the theoretical model based on experimentally obtained visibilities. (b) The orange curve corresponds to the Fisher information, with shaded regions indicating a 95\% confidence level, derived from the standard deviation in the fitting parameter. Purple and cyan dashed lines represent the SQL and the HS, respectively. (c) Data points are obtained via maximum likelihood estimation, with error bars derived using the bootstrapping method~\cite{slussarenko2017}. See Supplemental Material for the detailed calculation on error bars of standard deviation.~\cite{supple}}
\label{fig:figure3}
\end{figure}
	
To realize distributed quantum sensing, we performed the experiments with four-mode $2002$ states and $\bm{\nu}=(1/2, 1/2)$ for estimating the average of the unknown two phases $\phi=(\phi_1+\phi_2)/2$. The detection probabilities, experimentally obtained by simultaneously scanning $\phi_1$ and $\phi_2$ while maintaining $\phi_1=\phi_2$, are shown in Fig.~\ref{fig:figure3}(a). The visibilities of the interference fringes for $\{P_{1}^{11}$, $P_{1}^{20}$, $P_{1}^{02}$, $P_{2}^{11}$, $P_{2}^{20}$, $P_{2}^{02}\}$ were obtained to be $0.97^{+0.01}_{-0.01},\;0.96^{+0.01}_{-0.01},\;0.96^{+0.01}_{-0.01},\;0.94^{+0.01}_{-0.01},\;0.97^{+0.01}_{-0.01}$, and $0.99^{+0.00}_{-0.01}$, respectively. To verify the quantum enhancement, we use the Fisher information, defined as $1/(\nu^T \bold{F}_\text{C}^{-1} \nu)$, which is obtained from the FIM based on experimentally obtained two-photon detection probabilities, as shown in Fig.~\ref{fig:figure3}(b)~\cite{liu2021,zhao2021}. See Supplemental Material for the experimentally obtained FIM, $\bold{F}_\text{C(\text{exp})}$~\cite{supple}. The experimentally obtained maximum Fisher information is $3.76^{+0.03}_{-0.02}$ beyond the SQL of 2 as shown in Fig.~\ref{fig:figure3}(b). Then, an average of unknown two phases $\phi_\text{est}$ can be obtained using MLE, and the standard deviation is calculated as $\Delta \phi =\sqrt{\bm{\nu}^T \bold{F}_\text{C}^{-1} \bm{\nu}/\mu}$, as shown in Fig.~\ref{fig:figure3}(c). To this end, we employed a standard bootstrapping method~\cite{slussarenko2017}. The SQL and the HS were also drawn by $\Delta \phi_\text{SQL}=1/\sqrt{\mu N}$ and $\Delta \phi_\text{HS}=1/\sqrt{\mu}N$ with $\mu\simeq4931$, respectively.  See Supplemental Material for details on experimental estimation error~\cite{supple}.  In addition, we show that the standard deviation of estimated phases closely matches the HS, surpassing the SQL with a $2.74^{+0.04}_{-0.02}$ dB enhancement. Our experimental results in distributed quantum sensing explicitly demonstrate that the use of multi-mode $N00N$ states can achieve the HS of 1/$N^2$.  Note that we used post-selection, which does not take into account the imperfections of the experiment; however, this does not affect the proof-of-concept of quantum-enhanced sensitivity.

In summary, we theoretically identified the sensitivity bounds of multi-mode $N00N$ states for estimating spatially distributed multiple parameters. In particular, the QCRB of multi-mode $N00N$ states can provide quantum-enhanced sensitivity equivalent to the HS. Additionally, the CRB of multi-mode $N00N$ states can reach the QCRB with local measurements. Furthermore, we experimentally demonstrated distributed quantum sensing for estimating the average of two unknown phases, achieving 2.74 dB quantum-enhanced sensitivity over the SQL using four-mode $2002$ states. 
	
While we have focused on estimating two unknown phases using four-mode $2002$ states, these states can be extended to higher-mode $N00N$ states by generating the proposed superposition of the Bell state and using a double Sagnac interferometer~\cite{kim2024,hu2018}, enabling the estimation of more unknown phases. The use of higher-mode $N00N$ states allows for the estimation of an average of unknown phases even when the number of photons is less than the number of phases~\cite{kim2024}. Furthermore, a recent proposal has reported the possibility of generating high-photon-number multi-mode $N00N$ states using metasurfaces, suggesting a feasible approach for scaling up the photon number~\cite{tian2024}. Another intriguing direction is to explore arbitrary weighted linear global functions of unknown phases by employing phase shifters multiple times~\cite{liu2021,zhao2021}. Finally, investigating methods to achieve enhanced sensitivity beyond the SQL in the presence of decoherence is another promising direction~\cite{namkung2024,namkung20242,dorner2009, kacprowicz2010}. Our results can contribute to the theoretical understanding of quantum-enhanced multiple-parameter estimation and advance distributed quantum sensing for large-scale practical applications.
\\

This work was partly supported by National Research Foundation of Korea (NRF) grant funded by the Korea government (MSIT) (RS-2022-NR068817, 2023M3K5A1094805, RS-2024-00336079), Institute for Information \& communications Technology Planning\&Evaluation (IITP) grant funded by the Korea government (MSIT) (RS-2023-00222863), the National Research Council of Science and Technology (NST) by the Korea government (MSIT) (CAP21034-000), National Information Society Agency (NIA) funded by the Korea government (MSIT) [Quantum technology test verification and consulting support in 2024], and the KIST research program (2E33541, 2E33571).

\end{document}